\documentclass[aps,prd,twocolumn,nofootinbib,superscriptaddress]{revtex4-2}
\usepackage{amsmath,amssymb}
\usepackage{graphicx}
\usepackage{placeins} 
\usepackage[colorlinks=true,allcolors=blue]{hyperref}

\usepackage{orcidlink}
\begin{document}

\title{Axion-angle dependence of the rotational and tidal properties of PNJL quark stars}

\author{Safiqul Islam \orcidlink{0000-0003-1373-4137}}
\email{Corresponding author:sislam@kfu.edu.sa}
\affiliation{Department of Mathematics and Statistics, College of Science,
King Faisal University, P.O. Box 400, Al Ahsa 31982, Saudi Arabia}

\author{Ayan Banerjee \orcidlink{0000-0003-3422-8233}}
\email{ayanbanerjeemath@gmail.com}
\affiliation{Astrophysics and Cosmology Research Unit, School of Mathematics, Statistics and Computer Science, University of KwaZulu--Natal, Private Bag X54001, Durban 4000, South Africa}

\author{Javlon~Rayimbaev\orcidlink{0000-0001-9293-1838}}
\email{javlon@astrin.uz}
\affiliation{Institute of Theoretical Physics, National University of Uzbekistan, Tashkent 100174, Uzbekistan}
\affiliation{University of Tashkent for Applied Sciences, Str. Gavhar 1, Tashkent 100149, Uzbekistan}
\affiliation{Tashkent State Technical University, Tashkent 100095, Uzbekistan}

\author{Safyan Mukhtar \orcidlink{0000-0001-5415-6762}}
\email{smahmad@kfu.edu.sa}
\affiliation{Department of Mathematics and Statistics, College of Science,
King Faisal University, P.O. Box 400, Al Ahsa 31982, Saudi Arabia}

\date{\today}

\begin{abstract}
We investigate how a fixed, spatially uniform axion angle modifies the rotational and static tidal properties of self-bound quark stars in general relativity, using an axion-extended Polyakov-loop Nambu--Jona-Lasinio equation of state.
We vary $\theta$ from $0$ to $\pi$ while holding the remaining model parameters fixed, and compute the leading-order moment of inertia and the second-order spin-induced quadrupole moment and mass shift.
The calculation includes finite-surface-density contributions to the rotational mass shift and static tidal response.
Across this range, the static maximum mass increases from $1.68$ to $1.95\,M_\odot$, remaining below the lower edge of the quoted $1\sigma$ timing interval for PSR~J0740+6620.
At $M=1.338\,M_\odot$, the moment of inertia increases by $13\%$ and the dimensionless tidal deformability rises from $342$ to $547$.
Over the common mass range examined, the differences in the moment of inertia between adjacent sampled angles exceed the tested interpolation and table-resolution sensitivities.
Despite these changes, joint fits to the five sequences describe the $\bar I$--$C$ and I-Love-Q relations with maximum departures of $2.7\%$ and $0.4\%$, respectively, over the fitted domains.
The predicted binary tidal-deformability curves lie within the estimated 90\% credible region of the adopted GW170817 posterior over the mass-ratio ranges examined.
Within this EOS family, stellar properties vary appreciably with $\theta$, while the normalized quantities admit accurate joint fits across the five sequences.
\end{abstract}

\maketitle


\section{Introduction}
\label{sec:intro}

The rotational and tidal responses of compact stars provide complementary probes of dense matter beyond the mass--radius relation.
Rotation introduces the angular momentum, the spin-induced quadrupole moment $Q$ and a correction to the gravitational mass; the moment of inertia $I$ characterizes the leading response of the angular momentum to the stellar spin.
The moment of inertia may be measured through spin--orbit coupling in double neutron-star binaries~\cite{Lattimer:2005constraining}.
The quadrupole moment affects the X-ray pulse profiles of spinning stars~\cite{Psaltis:2014pulse} and the gravitational waves of binary inspirals~\cite{Ryan:1997accuracy,Yagi:2013iloveq}.
It also shapes the exterior spacetime used in models of quasi-periodic oscillations~\cite{Pappas:2012what,Boshkayev:2012gravitational}.
The exterior gravitational field can be characterized through the multipole moments of a stationary spacetime~\cite{Geroch:1970multipole,Thorne:1980multipole,VanVoorthuizen:2025multipole}.

Most observed pulsars spin well below their mass-shedding limit.
For them, rotation can be treated as a small perturbation of a static star.
This slow-rotation expansion was developed by Hartle~\cite{Hartle:1967slowly} and by Hartle and Thorne~\cite{Hartle:1968slowly}, to second order in the angular velocity $\Omega$.
Hartle later computed the rotational correction to the moment of inertia~\cite{Hartle:1973slowly}.
Comparisons with fully rotating models show that the second-order results are accurate at the spin rates of most observed pulsars~\cite{Berti:2004approximate,Berti:2005rotatinga,Benhar:2005perturbative,Yagi:2014effective}.
Rapidly rotating stars require full numerical solutions~\cite{Komatsu:1989rapidly,Cook:1994rapidly,Cook:1994rapidlya,Stergioulas:1995comparing,Paschalidis:2017rotating}.
The slow-rotation approach is still being developed, for example for differential rotation~\cite{Chirenti:2013slowly} and for high-accuracy modelling~\cite{Beltracchi:2024slowly,Conde-Ocazionez:2025highlyaccurate}.

Tidal interactions provide a complementary probe of stellar structure: during a binary inspiral, each star is deformed by the gravitational field of its companion.
The response is measured by the dimensionless tidal deformability $\bar\Lambda$, which enters the gravitational-wave phase~\cite{Flanagan:2008constraining,Hinderer:2008tidal,Hinderer:2010tidal,Damour:2009relativistic}.
In the observational comparison of Fig.~\ref{fig:gw}, $\Lambda_1$ and $\Lambda_2$ denote this same dimensionless quantity for the two stars.
Its measurability has been studied in detail~\cite{Hotokezaka:2016measurability}, and GW170817 has placed the first bounds on it~\cite{Abbott:2017gw170817,Abbott:2018gw170817,Abbott:2019properties}.
Independent observations constrain the masses and radii that viable stellar models must accommodate.
NICER has measured the masses and radii of several pulsars~\cite{Riley:2019nicer,Miller:2019psr,Riley:2021nicer,Miller:2021radius,Choudhury:2024nicer}.
Radio timing has found pulsars of about $2\,M_\odot$~\cite{Demorest:2010twosolarmass,Antoniadis:2013massive,Cromartie:2019relativistic,Fonseca:2021refined}, and GW190814 contained a compact object of about $2.6\,M_\odot$~\cite{Abbott:2020gw190814}.

The rotational and tidal responses are also connected through approximate relations among dimensionless stellar properties.
For many equations of state (EOSs), the normalized quantities $\bar I$, $\bar\Lambda$ and $\bar Q$ follow approximate EOS-insensitive relations, the I-Love-Q relations~\cite{Yagi:2013iloveqa,Yagi:2013iloveq}.
These relations hold to about 1\% for a wide range of EOSs.
They have been extended to anisotropic stars~\cite{Yagi:2015iloveqa} and are reviewed in Ref.~\cite{Yagi:2017approximate}.
The relation between $\bar I$ and the compactness $C$ is less tight~\cite{Lattimer:2005constraining}.

Self-bound quark stars are a distinct case.
If strange quark matter is the true ground state of baryonic matter, compact stars may be made entirely of it~\cite{Bodmer:1971collapsed,Witten:1984cosmica,Alcock:1986strangea}.
Such stars are reviewed in Refs.~\cite{Weber:2005strangea,Xia:2025strange}, and they remain an active topic~\cite{Teruya:2026selfbound}.
They are bound by the strong interaction rather than by gravity, and their energy density stays finite at the surface.
This density jump adds surface terms to the rotational mass shift and to the tidal Love number~\cite{Damour:2009relativistic,Postnikov:2010tidal,Reina:2015revisiting}.
The rotational properties of strange stars have been studied before~\cite{Miller:1991rotational,Urbanec:2013quadrupole,Yagi:2014effective}.
Their tidal deformability has also been computed, including the effect of repulsive vector interactions~\cite{Postnikov:2010tidal,Albino:2021tidal}.
Quark stars follow the I-Love-Q relations as well~\cite{Yagi:2013iloveqa,Yagi:2017approximate}.

A separate question concerns the QCD vacuum angle $\theta$.
A nonzero $\theta$ violates CP symmetry, and the axion was proposed to relax it dynamically.
In hot and dense matter, the axion potential and the topological susceptibility change~\cite{Zhang:2023exploring,Murgana:2024topological,Gong:2024qcd}.
In quark matter, $\theta$ also changes the chiral condensates and hence the EOS.
These effects have been studied with Nambu--Jona-Lasinio (NJL) type models~\cite{Boomsma:2009high,Sasaki:2012theta,Lu:2019effect,Abhishek:2021medium,Zhang:2025term}.
Their consequences have been explored for hybrid stars~\cite{Lopes:2022axion,Liu:2025axion,Kumar:2026cp}, for quark stars~\cite{Chu:2024properties} and for compact-star matter~\cite{Lu:2025qcd}.

Zhou \textit{et al.}~\cite{Zhou:2026effects} built an axion-extended Polyakov-loop NJL (PNJL) EOS for self-bound quark stars.
They treated $\theta$ as a fixed parameter in the range $0\le\theta\le\pi$.
They computed static sequences in general relativity (GR) and in four-dimensional Einstein--Gauss--Bonnet gravity, and found that a larger $\theta$ gives heavier stars.
A different line of work treats the axion as a field sourced inside the star, which changes the tidal relations directly~\cite{Kahn:2025effects}.
That mechanism differs from a fixed $\theta$ in the quark-matter EOS, which is the case we consider.

To our knowledge, the rotational properties of these stars have not been computed as functions of $\theta$. 
In this paper we compute them for the EOS of Ref.~\cite{Zhou:2026effects}, in GR.
We treat $\theta$ as a fixed, spatially uniform parameter and hold the remaining EOS parameters constant.
We obtain the leading-order moment of inertia and the quadrupole moment and mass shift at second order in $\Omega$.
We also compute the static tidal deformability of the same models, so that the I-Love-Q relations can be tested.
All self-bound surface terms are included.
Our aim is to quantify the dependence of these observables on $\theta$ at fixed mass and to assess how accurately joint fits describe the normalized quantities across the five sequences.
This comparison separates changes in individual stellar observables from departures from a joint description of the five EOS sequences.

We find that $\theta$ changes the moment of inertia at fixed mass.
At $M=1.338\,M_\odot$, $I$ grows from $1.238$ to $1.401\times10^{45}$~g~cm$^2$ between $\theta=0$ and $\pi$.
Over the common mass range examined, the minimum separation between neighbouring $\theta$ sequences exceeds the maximum tested interpolation sensitivity by a factor of about 12 for the closest pair, and by more for the others.
In contrast, the normalized relations stay close to common fits.
The $\bar I$--$C$ relation departs from its fit by at most $2.7\%$, and the I-Love-Q relations by at most $0.4\%$, over their fitted domains.
None of the static sequences reaches $2\,M_\odot$.

The paper is organized as follows.
Section~\ref{sec:eos} describes the EOS.
Section~\ref{sec:method} gives the slow-rotation and tidal equations and our numerical checks.
Section~\ref{sec:results} presents the results, and Sec.~\ref{sec:conclusions} concludes.
We use $G=c=1$ unless units are shown.


\section{Axion-PNJL quark matter}
\label{sec:eos}

\subsection{Model}

We use the $(2+1)$-flavour PNJL model with axion interactions of Zhou \textit{et al.}~\cite{Zhou:2026effects}.
It extends the SU(3) NJL and PNJL models~\cite{Rehberg:1996hadronization,Rossner:2007polyakov} by a $\theta$-dependent 't~Hooft term~\cite{Boomsma:2009high,Sasaki:2012theta,Lopes:2022axion,Chu:2024properties} and by a vector interaction~\cite{Shao:2013isoscalarvector}.
The Lagrangian density is
\begin{multline}
  \mathcal L=\bar\psi\left(i\gamma^\mu D_\mu+\gamma^0\hat\mu-\hat m\right)\psi
  +\mathcal L_S+\mathcal L_V\\
  -K\left\{e^{i\theta}\det{}_f\!\left[\bar\psi(1+\gamma_5)\psi\right]
  +e^{-i\theta}\det{}_f\!\left[\bar\psi(1-\gamma_5)\psi\right]\right\}\\
  -\mathcal U(\Phi,\bar\Phi,T),
  \label{eq:lagrangian}
\end{multline}
with the scalar and vector interactions
\begin{align}
  \mathcal L_S&=G_S\sum_{a=0}^{8}\left[(\bar\psi\lambda_a\psi)^2+(\bar\psi i\gamma_5\lambda_a\psi)^2\right],
  \label{eq:LS}\\
  \mathcal L_V&=-G_V\sum_{a=0}^{8}\left[(\bar\psi\gamma^\mu\lambda_a\psi)^2+(\bar\psi\gamma^\mu\gamma_5\lambda_a\psi)^2\right].
  \label{eq:LV}
\end{align}
Here $\psi$ is the quark field with flavours $u$, $d$ and $s$, $\hat m=\mathrm{diag}(m_u,m_d,m_s)$ is the current mass matrix and $\hat\mu$ is the matrix of quark chemical potentials.
The covariant derivative $D_\mu=\partial_\mu-iA_\mu$ contains a static, homogeneous gluon background field, which enters through the traced Polyakov loop $\Phi$ and its conjugate $\bar\Phi$.
$\mathcal U$ is the Polyakov-loop potential of Ref.~\cite{Rossner:2007polyakov}.

The six-quark 't~Hooft term carries the axion angle through the phases $e^{\pm i\theta}$, with $\theta=a/f_a$ for an axion field $a$ and decay constant $f_a$.
We treat $\theta$ as a fixed, uniform background, as in Ref.~\cite{Zhou:2026effects}.
The interaction is $2\pi$-periodic in $\theta$, and the thermodynamics is even under $\theta\to-\theta$, so $0\le\theta\le\pi$ covers all independent values.
We use $\theta=0,\ \pi/3,\ \pi/2,\ 2\pi/3$ and $\pi$.

A nonzero $\theta$ mixes the scalar and pseudoscalar channels~\cite{Boomsma:2009high,Sasaki:2012theta}.
In the mean-field approximation, each flavour has a scalar condensate $\sigma_i=-\langle\bar\psi_i\psi_i\rangle$ and a pseudoscalar condensate $\eta_i=-\langle\bar\psi_i i\gamma_5\psi_i\rangle$.
The constituent mass is $M_i=[(M_i^s)^2+(M_i^p)^2]^{1/2}$, with
\begin{align}
  M_i^s&=m_i-4G_S\sigma_i+2K\cos\theta\,(\sigma_j\sigma_k-\eta_j\eta_k)\nonumber\\
  &\quad+2K\sin\theta\,(\sigma_j\eta_k+\sigma_k\eta_j),
  \label{eq:Ms}\\
  M_i^p&=4G_S\eta_i+2K\cos\theta\,(\sigma_j\eta_k+\sigma_k\eta_j)\nonumber\\
  &\quad-2K\sin\theta\,(\sigma_j\sigma_k-\eta_j\eta_k),
  \label{eq:Mp}
\end{align}
where $(i,j,k)$ is a cyclic permutation of $(u,d,s)$.
For $\theta=0$ the pseudoscalar condensates vanish and Eq.~\eqref{eq:Ms} reduces to the usual NJL mass.

\subsection{Equation of state}

For the stellar configurations considered here, we use the zero-temperature limit.
At $T=0$ the Polyakov-loop potential vanishes and the quark distributions become step functions, so the Polyakov loop decouples.
The model then reduces to an axion-NJL model with vector interactions.
The thermodynamic potential is
\begin{align}
  \Omega&=\Omega_q+2G_S\sum_i\left(\sigma_i^2+\eta_i^2\right)-2G_V\sum_i\rho_i^2\nonumber\\
  &\quad-4K\cos\theta\left(\sigma_u\sigma_d\sigma_s-\sigma_u\eta_d\eta_s-\eta_u\sigma_d\eta_s-\eta_u\eta_d\sigma_s\right)\nonumber\\
  &\quad+4K\sin\theta\left(\eta_u\eta_d\eta_s-\eta_u\sigma_d\sigma_s-\sigma_u\eta_d\sigma_s-\sigma_u\sigma_d\eta_s\right)\nonumber\\
  &\quad+\Omega_\ell,
  \label{eq:omega}
\end{align}
with the quark contribution
\begin{multline}
  \Omega_q=-2N_c\sum_i\int\!\frac{d^3p}{(2\pi)^3}\Big[E_i\,\Theta(\Lambda_{\rm NJL}-p)\\
  +(\tilde\mu_i-E_i)\,\Theta(\tilde\mu_i-E_i)\Big].
  \label{eq:omegaq}
\end{multline}
Here $E_i=(p^2+M_i^2)^{1/2}$, $N_c=3$ and $\Lambda_{\rm NJL}$ is the momentum cutoff.
The vector interaction shifts the chemical potentials to $\tilde\mu_i=\mu_i-4G_V\rho_i$, where $\rho_i=N_c(\tilde\mu_i^2-M_i^2)^{3/2}/(3\pi^2)$ is the quark number density (zero for $\tilde\mu_i<M_i$).
$\Omega_\ell$ is the contribution of electrons and muons, treated as free Fermi gases.
The condensates follow from the gap equations
\begin{equation}
  \frac{\partial\Omega}{\partial\sigma_i}=\frac{\partial\Omega}{\partial\eta_i}=0 .
  \label{eq:gap}
\end{equation}

Stellar matter is electrically neutral and in $\beta$ equilibrium,
\begin{gather}
  \mu_d=\mu_s=\mu_u+\mu_e,\qquad \mu_\mu=\mu_e,\nonumber\\
  \tfrac23\rho_u-\tfrac13\rho_d-\tfrac13\rho_s-\rho_e-\rho_\mu=0 .
  \label{eq:beta}
\end{gather}
The pressure and energy density are
\begin{equation}
  P=-(\Omega-\Omega_0)-B,\qquad
  \varepsilon=-P+\sum_{f}\mu_f\rho_f ,
  \label{eq:Peps}
\end{equation}
where the sum runs over quarks and leptons and $\Omega_0$ is the vacuum value of $\Omega$.
The bag constant $B$ makes the matter self-bound: the pressure vanishes at a finite energy density.
The baryon density is $\rho_B=(\rho_u+\rho_d+\rho_s)/3$, and $\rho_0=0.16$~fm$^{-3}$.

We use the parameters of Ref.~\cite{Zhou:2026effects}.
These are $\Lambda_{\rm NJL}=630$~MeV, $m_u=m_d=5.5$~MeV, $m_s=135.7$~MeV, $G_S\Lambda_{\rm NJL}^2=1.781$ and $K\Lambda_{\rm NJL}^5=9.29$~\cite{Rehberg:1996hadronization}.
The vector coupling is $G_V=0.5\,G_S$~\cite{Shao:2013isoscalarvector,Chu:2024properties}, and $B^{1/4}=93$~MeV~\cite{Yuan:2026nonstrange}.
Only $\theta$ is varied.

\subsection{Properties of the EOS}

We retain the original tabulated EOS values for each \(\theta\) and restrict the calculations to the tabulated pressure range; interpolation between nodes is described in Sec.~\ref{sec:numerics}.
Each table starts at $P=0$, where the surface energy density is $\varepsilon_s\simeq230$--$239$~MeV~fm$^{-3}$.

\begin{figure*}[tp]
  \centering
  \includegraphics[width=\textwidth]{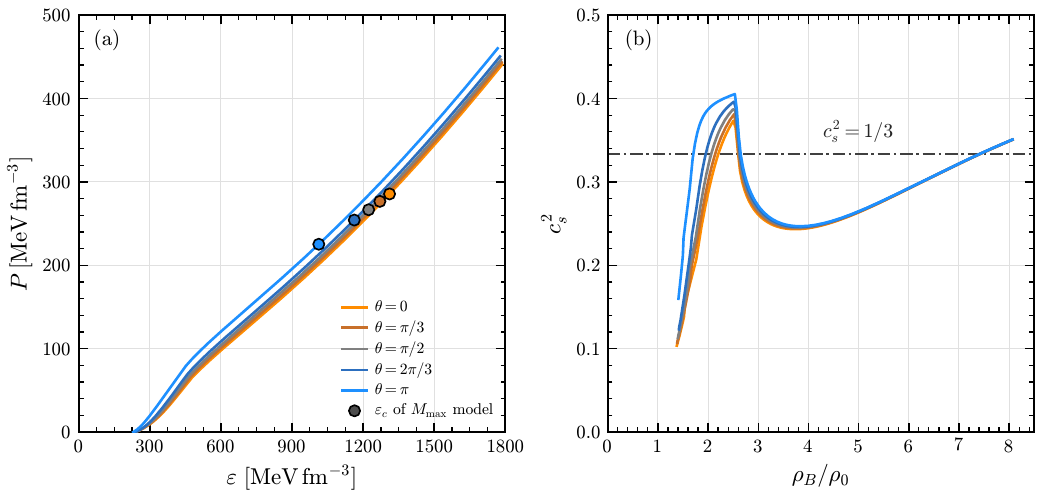}
  \caption{Axion-PNJL EOS of Ref.~\cite{Zhou:2026effects} for five values of $\theta$, over the full table range.
  (a) Pressure against energy density; circles mark the central energy density of each maximum-mass star.
  (b) Squared sound speed at the table nodes against baryon density.
  The dash-dotted line is $c_s^2=1/3$.}
  \label{fig:eos}
\end{figure*}

Figure~\ref{fig:eos} shows the EOS.
At fixed $\varepsilon$, the pressure increases with $\theta$ above $\varepsilon\approx250$~MeV~fm$^{-3}$, so $\theta=\pi$ gives the stiffest EOS.
The effect is largest at low density: at $\varepsilon=300$~MeV~fm$^{-3}$ the pressure for $\theta=\pi$ is about twice that for $\theta=0$, while at $1500$~MeV~fm$^{-3}$ the difference is $7\%$.
The squared sound speed $c_s^2=dP/d\varepsilon$ peaks near $\rho_B\approx2.5\rho_0$, at $0.373$--$0.405$.
It exceeds the conformal value $1/3$ around this peak, from $\rho_B\approx1.7\rho_0$ for $\theta=\pi$ and $2.2\rho_0$ for $\theta=0$.
Zhou \textit{et al.} attribute the fall above the peak to the onset of strange quarks~\cite{Zhou:2026effects}; a similar peak appears without $\beta$ equilibrium~\cite{He:2022speed}.
The sound speed exceeds $1/3$ again only above $\rho_B\approx7.4\rho_0$, a region that no stable star reaches.
At every table node $c_s^2<1$.

\section{Slowly rotating stars in general relativity}
\label{sec:method}

We describe rotation with the Hartle--Thorne expansion to second order in the angular velocity $\Omega$~\cite{Hartle:1967variational,Hartle:1967slowly,Hartle:1968slowly}.
Tides are treated on the static background~\cite{Hinderer:2008tidal,Hinderer:2009erratum,Damour:2009relativistic}.
Lengths are given in km.
Only the equations needed to define our quantities are shown here; the full derivations are in the references above.

\subsection{Equilibrium and slow rotation}
\label{sec:equilibrium-rotation}

The static star has the metric
\begin{equation}
  ds^2=-e^{\nu}dt^2+e^{\lambda}dr^2+r^2\left(d\vartheta^2+\sin^2\vartheta\,d\phi^2\right),
  \label{eq:metric0}
\end{equation}
Here $\nu(r)$ and $\lambda(r)$ are the temporal and radial metric functions, respectively, with $e^{-\lambda}=1-2m(r)/r$ and $m(r)$ the enclosed gravitational mass.
The static configuration obeys the Tolman--Oppenheimer--Volkoff equations~\cite{Tolman:1939statica,Oppenheimer:1939massivea}.
We denote the energy density by $\varepsilon$, the pressure by $P$, and differentiation with respect to $r$ by a prime.
The enclosed mass satisfies
\begin{equation}
  m'=4\pi r^2\varepsilon,
  \label{eq:mass-background}
\end{equation}
while the temporal metric function is determined by
\begin{equation}
  \nu'=\frac{2(m+4\pi r^3P)}{r(r-2m)}.
  \label{eq:metric-gradient}
\end{equation}
Hydrostatic equilibrium relates the pressure gradient to this metric function through
\begin{equation}
  P'=-\tfrac12(\varepsilon+P)\,\nu' .
  \label{eq:tov}
\end{equation}
The stellar surface is located at $P(R)=0$, and the total gravitational mass is $M=m(R)$.

For a self-bound star, the energy density remains finite at this surface, dropping from $\varepsilon_s=\varepsilon(P=0)$ to zero outside the star.
The interior background is still governed by Eqs.~\eqref{eq:mass-background}--\eqref{eq:tov}.
The surface density jump requires additional matching contributions to the second-order rotational mass shift and the tidal response.
We stop the interior integration at $R$ and include these contributions analytically [Eqs.~\eqref{eq:m0jump} and~\eqref{eq:yjump}].

Along each fixed-$\theta$ sequence, configurations are parametrized by the central energy density $\varepsilon_c$.
Models beyond the maximum of $M(\varepsilon_c)$ are labelled unstable by the turning-point criterion.
We do not compute radial modes.

At first order, rotation adds the term $-2\omega(r)\,r^2\sin^2\vartheta\,dt\,d\phi$ to the metric.
Here $\omega(r)$ describes frame dragging, and $\Omega$ is the uniform angular velocity of the star. The relative angular velocity $\varpi=\Omega-\omega$ satisfies~\cite{Hartle:1967slowly}
\begin{equation}
  \varpi''=-\frac{4}{r}\left[1-\pi r^2e^{\lambda}(\varepsilon+P)\right]\varpi'
  +16\pi e^{\lambda}(\varepsilon+P)\,\varpi .
  \label{eq:framedrag}
\end{equation}
At the centre, the regular solution has a finite relative angular velocity and a vanishing radial derivative; we initialize it from a central series at a small positive radius, with unit central amplitude before normalization.
Outside the star $\varpi=\Omega-2J/r^3$, where $J$ is the angular momentum. The angular velocity is therefore obtained from $\Omega=\varpi(R)+2J/R^3$.
Continuity at the surface fixes the angular momentum and angular velocity of this solution. Matching at $R$ gives
\begin{equation}
  J=\frac{R^4\varpi'(R)}{6},\qquad I=\frac{J}{\Omega},\qquad \bar I=\frac{I}{M^3}.
  \label{eq:inertia}
\end{equation}
The ratio $I=J/\Omega$ characterizes the leading rotational response of the static configuration and is independent of the arbitrary amplitude of the regular first-order solution. At this order $I$ does not depend on $\Omega$.
Since $\varpi$ and $\varpi'$ are continuous at $R$, no additional surface-jump correction is required when extracting $J$ and $I$ from the matched first-order solution.

At second order the star is deformed.
The monopole perturbation $(m_0,p_0^*)$ changes the mass, and the quadrupole perturbation $(h_2,v_2)$ gives the quadrupole moment~\cite{Hartle:1968slowly}.
Here $p_0^*$ is the dimensionless monopole pressure-perturbation factor. We solve the equations of Refs.~\cite{Hartle:1967slowly,Hartle:1968slowly}, with $p_0^*(0)=0$.

We compare a rotating star with the static star of the same central energy density $\varepsilon_c$, at a fixed spin frequency.
The mass shift is then
\begin{equation}
  \delta M=m_0(R^-)+\frac{J^2}{R^3}+[m_0],
  \label{eq:deltaM}
\end{equation}
with the surface term
\begin{equation}
  [m_0]=\frac{4\pi R^3(R-2M)}{M}\,\varepsilon_s\,p_0^*(R).
  \label{eq:m0jump}
\end{equation}
The last term is the contribution of the surface density jump~\cite{Reina:2015revisiting}; $\delta M$ also enters extended universal relations~\cite{Reina:2017completion}.
The surface contribution $[m_0]$ vanishes when $\varepsilon_s=0$.
Since $\delta M\propto\Omega^2$, we report $\delta M/M$ at a given frequency.
The paired configurations are compared at fixed central energy density, rather than along the spin evolution of an individual star.
The rotating counterparts of the static maximum-mass configurations are therefore not identified as the maxima of rotating equilibrium sequences.

Matching $h_2$ and $v_2$ to the exterior solution fixes one constant $\mathcal K$, which gives the quadrupole moment~\cite{Hartle:1968slowly,Yagi:2013iloveq},
\begin{equation}
  Q=-\left(\frac{J^2}{M}+\frac{8}{5}\mathcal K M^3\right),\qquad
  \bar Q=-\frac{Q}{M^3\chi^2},
  \label{eq:quadrupole}
\end{equation}
with $\chi=J/M^2$.
For a Kerr black hole $\bar Q=1$.
Unlike $m_0$, the functions $h_2$ and $v_2$ are continuous at the surface, so $Q$ needs no additional surface term.

The equatorial radius $R_{\rm eq}$ is the circumferential radius of the deformed surface at the equator.
To second order,
\begin{equation}
  R_{\rm eq}=R+\xi_0-\tfrac12\left[\xi_2+R\,k_2^{\rm HT}(R)\right],
  \label{eq:Req}
\end{equation}
where $\xi_0$ and $\xi_2$ are the monopole and quadrupole displacements of the surface and $k_2^{\rm HT}=v_2-h_2$ is the angular metric function~\cite{Hartle:1968slowly}.
(The superscript separates it from the Love number $k_2$.)

The expansion requires $\Omega\ll\Omega_\star$, with $\Omega_\star=\sqrt{M/R^3}$.
At the spin frequency of PSR J0740+6620, $\nu\simeq346$~Hz~\cite{Cromartie:2019relativistic},
we show only models with $\Omega/\Omega_\star\le0.20$.
This is a practical selection, not an error bound.

\subsection{Static tidal response}
\label{sec:tides}

We denote by $H$ the radial metric-perturbation function describing the static quadrupolar ($\ell=2$) tidal response. Its logarithmic derivative $y=r H'/H$ obeys~\cite{Hinderer:2008tidal,Hinderer:2009erratum,Damour:2009relativistic},
\begin{multline}
  r y'+y^2+y\,e^{\lambda}\left[1+4\pi r^2(P-\varepsilon)\right]\\
  +4\pi r^2e^{\lambda}\left(5\varepsilon+9P+\frac{\varepsilon+P}{c_s^2}\right)
  -6e^{\lambda}-r^2\nu'^2=0,
  \label{eq:tidal}
\end{multline}
with $y(0)=2$ and $c_s^2=dP/d\varepsilon$.
The symbols $R^-$ and $R^+$ denote the interior and exterior limits at the stellar surface, respectively. At a self-bound surface $y$ jumps~\cite{Damour:2009relativistic,Postnikov:2010tidal},
\begin{equation}
  y(R^+)=y(R^-)-\frac{4\pi R^3\varepsilon_s}{M}.
  \label{eq:yjump}
\end{equation}
The Love number $k_2$ is calculated using the corrected exterior value $y(R^+)$ and the compactness $C=M/R$~\cite{Hinderer:2008tidal,Hinderer:2009erratum},
and the dimensionless tidal deformability is $\bar\Lambda=\tfrac23k_2C^{-5}$.

\subsection{Numerical implementation and validation}
\label{sec:numerics}

The equation of state is the axion-PNJL table of Ref.~\cite{Zhou:2026effects}, used without change.
We interpolate $\varepsilon(P)$ with a monotone piecewise-cubic Hermite (PCHIP) interpolant,
which keeps $\varepsilon$ and $c_s^2$ continuous.
We use the adaptive DOP853 integrator, initialized from series solutions near the centre.
It stops at every table node and at the surface, so that no step crosses a node or $R$.

We tested the code on a self-bound linear equation of state, $\varepsilon=3P+4B_0$ with $B_0=60$~MeV~fm$^{-3}$.
For 42 stars with $6.7\times10^{-4}\le C\le0.28$, the code recovers the expected low-compactness behaviour of $\bar I$, $\bar Q$, $\delta M$ and $k_2$.
The mass shift from exterior matching agrees with a separate Komar-integral expression to approximately $5\times10^{-14}$ in relative terms.
At two benchmark configurations, a 30-digit Taylor integration agrees to $7\times10^{-15}$ across $M$, $R$, $\bar I$, $\bar Q$, $\delta M/\Omega^2$ and $\bar\Lambda$.
A separately written implementation of the same specification agrees to $1.1\times10^{-13}$ across these six quantities on the 42-model grid.
These are maximum relative differences over the stated comparisons, rather than bounds on the physical error; the inter-code test does not exclude errors shared by the specification.

For the axion-PNJL table, changing the starting radius, the integrator and its tolerances changes the results by at most $3\times10^{-11}$ (relative).
Interpolation sensitivity is assessed using piecewise-linear interpolation of $\varepsilon(P)$ and cubic Hermite interpolation of the logarithm of the baryon chemical potential, with thermodynamic slopes at the nodes.
For the separate resolution test, we retain the dense low-pressure block up to the first pressure interval wider than $2$~MeV~fm$^{-3}$, then retain alternate nodes starting with the next node and always preserve the final table node.
These sensitivities are used in Sec.~\ref{sec:results-rotation}.
They characterize the numerical choices tested here; they do not quantify uncertainty in the EOS parameters or the truncation of the slow-rotation expansion.


\section{Results}
\label{sec:results}

All results use the five sequences $\theta=0,\ \pi/3,\ \pi/2,\ 2\pi/3$ and $\pi$.
In the sequence plots, solid lines denote stable static models and dashed lines denote computed models beyond the maximum mass; Fig.~\ref{fig:inertia}(b) uses the separate line conventions given in its caption.
The dashed lines end where the EOS table ends.

\subsection{Equilibrium and tidal properties}
\label{sec:results-static}

\begin{figure*}[tp]
  \centering
  \includegraphics[width=\textwidth]{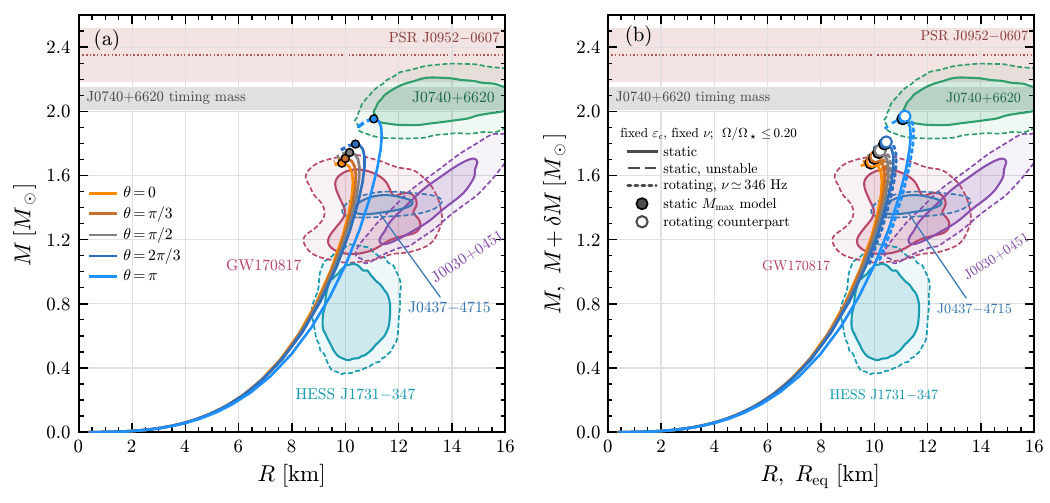}
  \caption{Mass--radius relations.
  (a) Static sequences; filled circles mark the maximum-mass models.
  (b) Static models and their rotating counterparts at $\nu\simeq346$~Hz (dotted) for $M\ge1\,M_\odot$; open circles mark the rotating counterparts of the static maximum-mass models.
  Shaded regions: NICER (68\%, 95\%; J0740+6620 cut at 16~km)~\cite{Miller:2019psr,Miller:2021radius,Choudhury:2024nicer}, GW170817 (50\%, 90\%)~\cite{Abbott:2018gw170817} and HESS~J1731$-$347 ($1\sigma$, $2\sigma$)~\cite{Doroshenko:2022strangely}.
  Horizontal bands: masses of PSR~J0740+6620~\cite{Fonseca:2021refined} and PSR~J0952$-$0607~\cite{Romani:2022psr}.}
  \label{fig:mr}
\end{figure*}

Figure~\ref{fig:mr}(a) shows the static mass--radius relations.
The maximum mass rises with $\theta$, from $1.677\,M_\odot$ at $\theta=0$ to $1.953\,M_\odot$ at $\theta=\pi$ (Table~\ref{tab:models}).
All five maximum masses lie below the lower edge of the quoted $1\sigma$ timing interval for PSR~J0740+6620, $2.08\pm0.07\,M_\odot$~\cite{Cromartie:2019relativistic,Fonseca:2021refined}.

\begin{table}[!t]
\centering
\caption{Model properties for the five values of $\theta$.
Top: the static maximum-mass model.
Middle: its rotating counterpart at $\nu=346$~Hz, at the same central energy density; $\Omega_\star=(M/R^3)^{1/2}$ and $R_{\rm eq}$ is the equatorial circumferential radius.
Bottom: static models at $M=1.338\,M_\odot$.
The top and bottom blocks list static background configurations and their associated rotational response coefficients; $\bar Q$ characterizes the spin-induced quadrupole response.
$\bar I=I/M^3$, $\bar Q=-Q/(M^3\chi^2)$ with $\chi=J/M^2$, and $\bar\Lambda=\tfrac23k_2C^{-5}$.}
\label{tab:models}
\setlength{\tabcolsep}{3pt}
\begin{tabular}{lccccc}
\hline\hline
$\theta$ & $0$ & $\pi/3$ & $\pi/2$ & $2\pi/3$ & $\pi$ \\
\hline
\multicolumn{6}{l}{\textit{Static maximum-mass configuration}} \\
$M_{\max}$ ($M_\odot$) & 1.6770 & 1.7063 & 1.7432 & 1.7957 & 1.9531 \\
$R$ (km) & 9.869 & 9.998 & 10.158 & 10.380 & 11.068 \\
$\varepsilon_c$ (MeV\,fm$^{-3}$) & 1311.5 & 1270.7 & 1223.3 & 1162.7 & 1013.0 \\
$C=M/R$ & 0.2509 & 0.2520 & 0.2534 & 0.2554 & 0.2606 \\
$k_2$ & 0.0622 & 0.0626 & 0.0629 & 0.0631 & 0.0625 \\
$\bar\Lambda$ & 41.69 & 41.03 & 40.11 & 38.70 & 34.68 \\
$I$ ($10^{45}$\,g\,cm$^2$) & 1.501 & 1.577 & 1.674 & 1.818 & 2.292 \\
$\bar I=I/M^3$ & 7.341 & 7.320 & 7.290 & 7.242 & 7.095 \\
$\bar Q$ & 2.746 & 2.732 & 2.712 & 2.681 & 2.591 \\
\hline
\multicolumn{6}{l}{\textit{Rotating counterpart at $\nu=346$\,Hz (same $\varepsilon_c$)}} \\
$\Omega/\Omega_\star$ & 0.1429 & 0.1444 & 0.1463 & 0.1489 & 0.1572 \\
$\delta M$ ($M_\odot$) & 0.0113 & 0.0119 & 0.0125 & 0.0134 & 0.0164 \\
$\delta M/M$ (\%) & 0.68 & 0.69 & 0.72 & 0.75 & 0.84 \\
$R_{\rm eq}-R$ (km) & 0.060 & 0.062 & 0.065 & 0.069 & 0.082 \\
\hline
\multicolumn{6}{l}{\textit{Static configuration at $M=1.338\,M_\odot$}} \\
$I$ ($10^{45}$\,g\,cm$^2$) & 1.2381 & 1.2550 & 1.2756 & 1.3041 & 1.4005 \\
$\bar I$ & 11.922 & 12.085 & 12.283 & 12.558 & 13.485 \\
$\bar Q$ & 5.242 & 5.329 & 5.434 & 5.580 & 6.069 \\
$\bar\Lambda$ & 342.01 & 360.49 & 383.86 & 417.86 & 547.18 \\
\hline\hline
\end{tabular}
\end{table}

\FloatBarrier

The figure also shows several observed sources as reference regions.
All five sequences intersect the displayed outer contours of PSR~J0437$-$4715~\cite{Choudhury:2024nicer}, GW170817~\cite{Abbott:2018gw170817} and HESS~J1731$-$347~\cite{Doroshenko:2022strangely}.
Only $\theta=\pi$ reaches the NICER region of PSR~J0740+6620~\cite{Miller:2021radius}, near its maximum mass.
This is possible because the 95\% NICER region extends below the $1\sigma$ timing band.
No sequence enters the region of PSR~J0030+0451~\cite{Miller:2019psr}.

The NICER radii are equatorial radii of spinning stars.
Figure~\ref{fig:mr}(b) adds the rotating counterpart of each static model at the spin of PSR~J0740+6620.
The same frequency is used for all sequences in this comparison.
Each rotating model has the same central energy density as its static model.
The mass rises by $\delta M$, and the equatorial radius exceeds the static radius by $0.06$--$0.17$~km.
The shift is small compared with the size of the observational regions.

\begin{figure*}[tp]
  \centering
  \includegraphics[width=\textwidth]{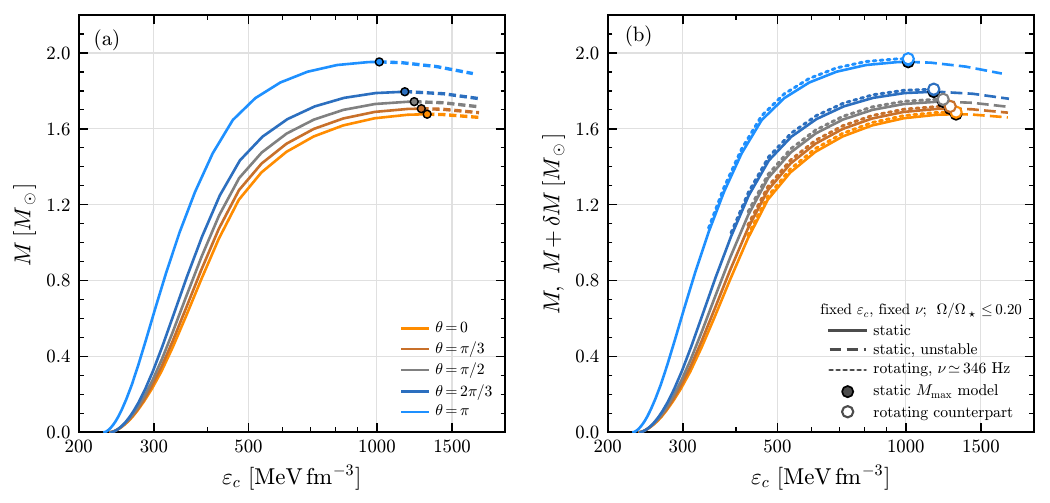}
  \caption{Gravitational mass against central energy density.
  (a) Static sequences.
  (b) Static models and their rotating counterparts at $\nu\simeq346$~Hz (dotted), at the same central energy density, for $M\ge1\,M_\odot$ and $\Omega/\Omega_\star\le0.20$.
  Filled circles mark the static maximum-mass models; open circles mark their rotating counterparts.}
  \label{fig:meps}
\end{figure*}

Figure~\ref{fig:meps} shows the same sequences against the central energy density.
The maximum masses occur at $\varepsilon_c=1312$~MeV~fm$^{-3}$ for $\theta=0$ and $1013$~MeV~fm$^{-3}$ for $\theta=\pi$.
These turning points define the stable branches used in all figures.
At fixed $\varepsilon_c$, each rotating model lies directly above its static model.
For the rotating counterparts of the static maximum-mass configurations, $\delta M=0.011$--$0.016\,M_\odot$.

\begin{figure*}[tp]
  \centering
  \includegraphics[width=\textwidth]{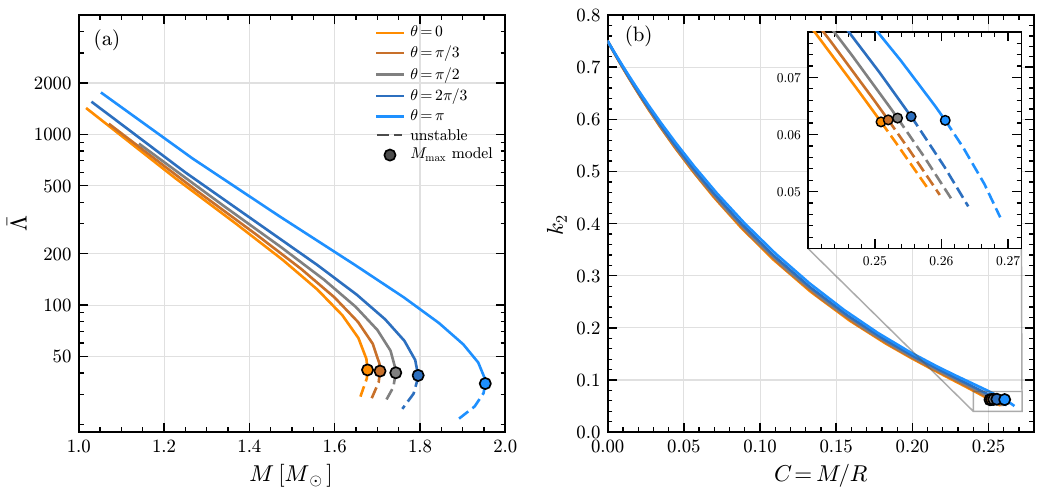}
  \caption{Static tidal properties.
  (a) Dimensionless tidal deformability $\bar\Lambda$ against mass, for $M\ge1\,M_\odot$.
  (b) Love number $k_2$ against compactness, including the stable branch and the computed continuation beyond the maximum mass; the inset enlarges the maximum-mass region.
  Circles mark the maximum-mass models.}
  \label{fig:tides}
\end{figure*}

Figure~\ref{fig:tides} shows the static tidal properties.
At fixed mass, $\bar\Lambda$ increases with $\theta$.
At $1.338\,M_\odot$ it rises from $342$ for $\theta=0$ to $547$ for $\theta=\pi$ (Table~\ref{tab:models}).
At the respective static maximum-mass configurations the order reverses: $\bar\Lambda$ falls from $41.7$ to $34.7$, because the $\theta=\pi$ star is more compact.
The lowest-compactness models give $k_2\simeq0.749$, approaching the homogeneous incompressible-star limit.
At fixed compactness, $k_2$ is ordered by $\theta$, with $\theta=\pi$ the largest.
The spread is $0.5\%$ at $C=0.01$ and grows to about $15\%$ near $C=0.24$.
At the maximum-mass models, however, $k_2$ stays within $0.0622$--$0.0631$, since each sequence ends at a different compactness.

\subsection{Rotational response}
\label{sec:results-rotation}

\begin{figure*}[tp]
  \centering
  \includegraphics[width=\textwidth]{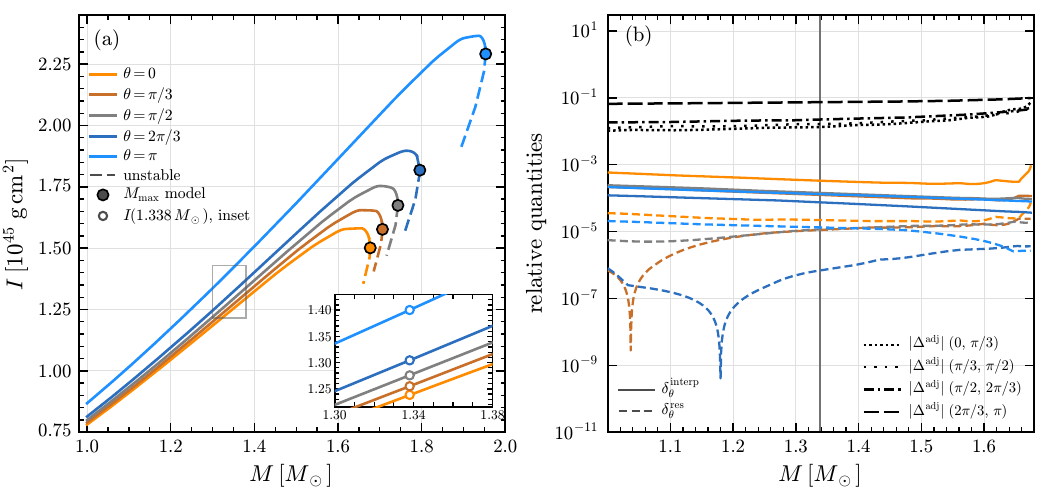}
  \caption{Moment of inertia.
  (a) $I$ against mass.
  The band around each curve (narrower than the line) is the spread between three EOS interpolants; the inset enlarges the region near $1.338\,M_\odot$.
  (b) Relative interpolation sensitivity $\delta^{\rm interp}_\theta$ (solid), resolution sensitivity $\delta^{\rm res}_\theta$ (dashed), and the relative difference $|\Delta^{\rm adj}|$ between neighbouring $\theta$ values (black dotted curves).
  The vertical line marks $1.338\,M_\odot$.}
  \label{fig:inertia}
\end{figure*}

Figure~\ref{fig:inertia}(a) shows the moment of inertia against mass.
At fixed mass, $I$ increases with $\theta$.
At $1.338\,M_\odot$, close to the mass of PSR~J0737$-$3039A in the double pulsar~\cite{Lattimer:2005constraining},
$I$ rises from $1.238$ to $1.401\times10^{45}$~g~cm$^2$ (Table~\ref{tab:models}), an increase of $13\%$.

\begin{figure*}[tp]
  \centering
  \includegraphics[width=\textwidth]{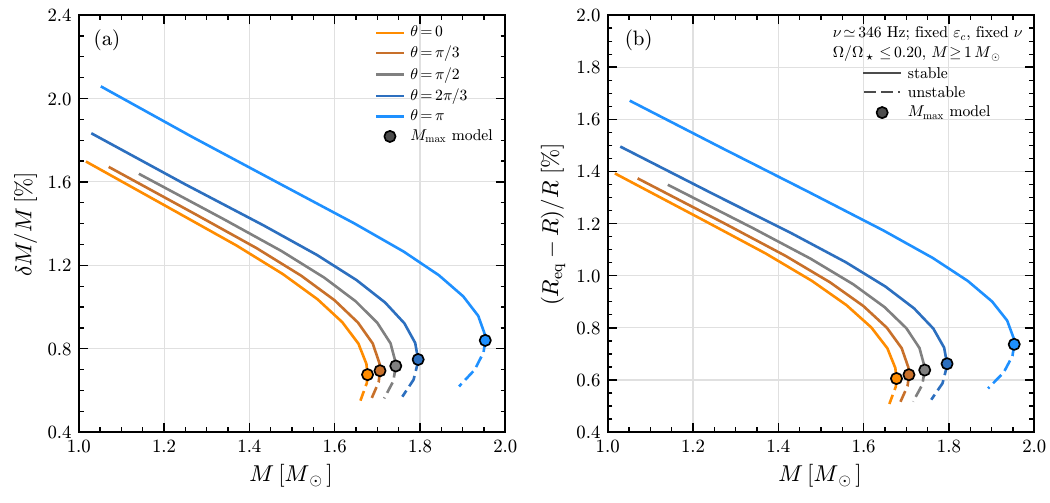}
  \caption{Second-order response at $\nu\simeq346$~Hz, at fixed central energy density.
  (a) Relative mass shift $\delta M/M$.
  (b) Fractional increase of the equatorial circumferential radius, $(R_{\rm eq}-R)/R$.
  Both are shown against the static mass, for $M\ge1\,M_\odot$ and $\Omega/\Omega_\star\le0.20$.}
  \label{fig:rotation}
\end{figure*}

\begin{figure*}[tp]
  \centering
  \includegraphics[width=\textwidth]{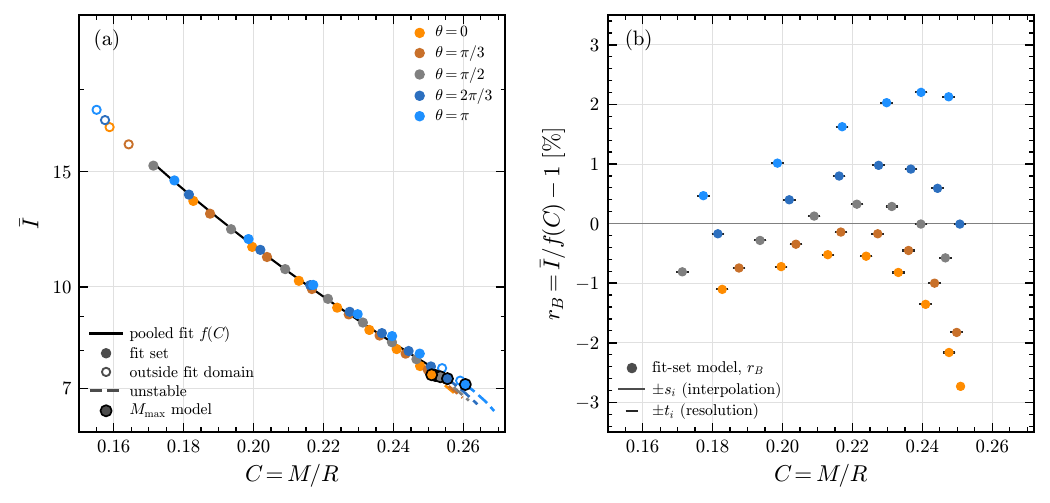}
  \caption{The $\bar I$--$C$ relation.
  (a) Stable models, dashed unstable continuations, and the common fit of Eq.~\eqref{eq:fit} (black line). Filled markers denote the fit set and open markers models outside it, except that the maximum-mass circles remain filled regardless of fit membership.
  (b) Relative departure from the fit.
  Bars show $s_i$ and caps $t_i$; both are smaller than the markers.}
  \label{fig:ibarc}
\end{figure*}

\begin{figure*}[tp]
  \centering
  \includegraphics[width=0.9\textwidth]{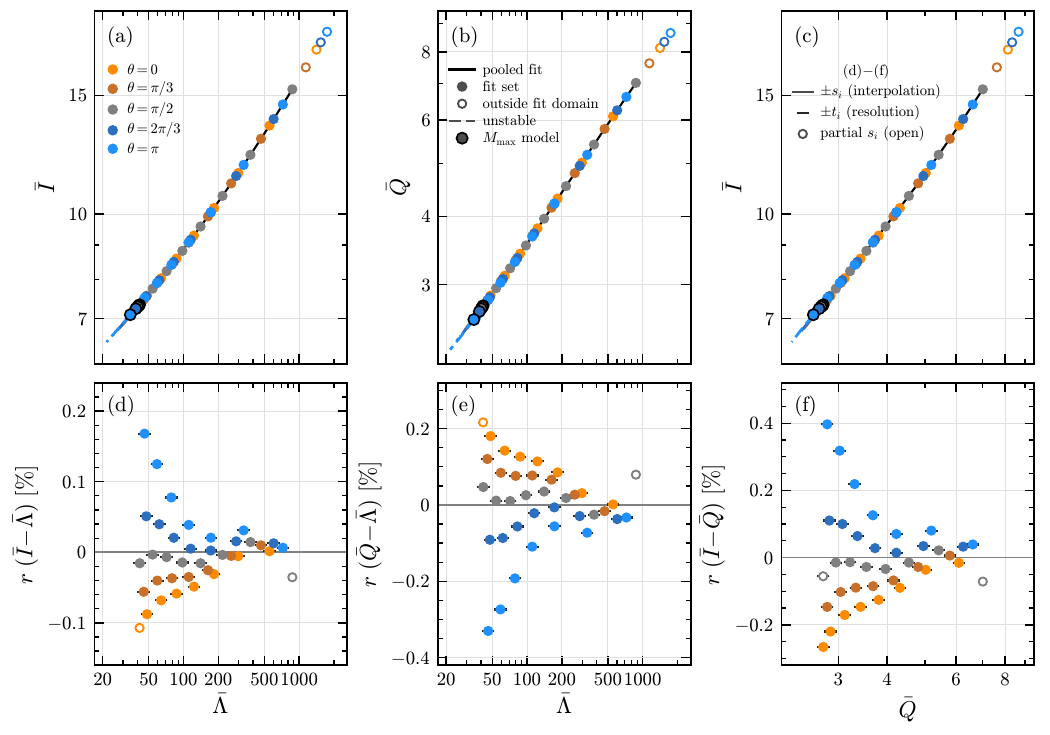}
  \caption{I-Love-Q relations.
  (a)--(c) $\bar I$--$\bar\Lambda$, $\bar Q$--$\bar\Lambda$ and $\bar I$--$\bar Q$, with common fits (black lines) drawn over their fit ranges and dashed unstable continuations. Filled markers denote the fit set and open markers models outside it; maximum-mass circles remain filled regardless of fit membership.
  (d)--(f) Relative departures from the fits, with $s_i$ and $t_i$ as in Fig.~\ref{fig:ibarc}; open circles mark models with only one interpolation variant.}
  \label{fig:ilq}
\end{figure*}

We test whether these differences are larger than our numerical sensitivities.
For each $\theta$ we recompute $I(M)$ with two other interpolants of the EOS table, and with a thinned table.
The largest relative changes define the interpolation and resolution sensitivities, $\delta^{\rm interp}_\theta$ and $\delta^{\rm res}_\theta$, shown in Fig.~\ref{fig:inertia}(b).
We compare them with the relative difference $|\Delta^{\rm adj}|$ in $I$ between neighbouring values of $\theta$, over the common mass range $1\le M/M_\odot\le1.674$.
The minimum separation exceeds the maximum interpolation sensitivity by a factor of about $11.8$ for $\theta=0$ and $\pi/3$; the factors for the other adjacent pairs are $51$, $75$ and $309$.
The resolution sensitivities are smaller still.

Figure~\ref{fig:rotation} shows the second-order response at $\nu\simeq346$~Hz.
The relative mass shift $\delta M/M$ falls with mass, from $1.6$--$2.1\%$ at the lowest masses shown to $0.68$--$0.84\%$ at the static maximum-mass configurations.
The fractional growth of the equatorial radius behaves in the same way, from $1.35$--$1.67\%$ to $0.61$--$0.74\%$.
Both increase with $\theta$ at fixed mass.
For all models shown, $\Omega/\Omega_\star$ lies between $0.14$ and $0.18$.

\subsection{Approximate relations within the EOS family}
\label{sec:results-relations}

\begin{figure*}[tp]
  \centering
  \includegraphics[width=0.9\textwidth]{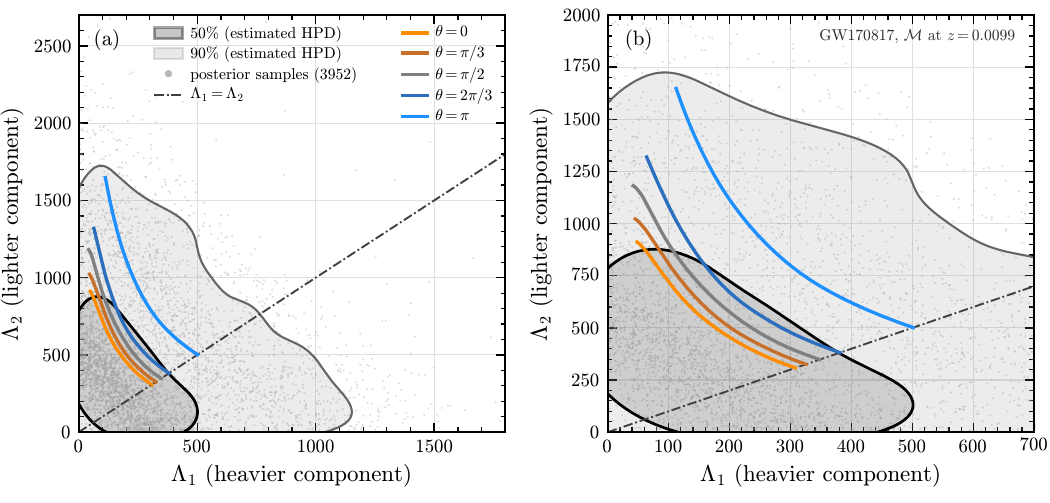}
  \caption{Tidal deformabilities of the two stars of GW170817, $\Lambda_1$ (heavier) and $\Lambda_2$ (lighter).
  (a) Full view; (b) enlargement.
  Grey dots are posterior samples~\cite{Abbott:2019properties}; the shaded regions are the 50\% and 90\% regions of a kernel-density estimate.
  Curves are static stable models at the source-frame chirp mass.
  The dash-dotted line is $\Lambda_1=\Lambda_2$.}
  \label{fig:gw}
\end{figure*}

Figure~\ref{fig:ibarc} shows $\bar I=I/M^3$ against the compactness.
We fit one curve to all five sequences,
\begin{equation}
  \ln\bar I=\sum_{k=0}^{n}a_k u^k,\qquad u=\frac{\ln C-\mu}{s},
  \label{eq:fit}
\end{equation}
where $\mu$ and $s$ centre and scale $\ln C$.
Each $\theta$ has equal weight, and the degree $n$ is chosen by leaving out one $\theta$ at a time.
The fit uses 35 stable models with $0.171\le C\le0.251$, and selects $n=1$.
The largest relative departure from the fit is $2.73\%$, and the $\theta$-balanced RMS logarithmic residual is $1.13\%$.
The departures are ordered by $\theta$ in Fig.~\ref{fig:ibarc}(b).
They also include the error of the fit itself.
For each model, $s_i$ measures the largest change in the residual under the other interpolants, and $t_i$ measures the change under table thinning.
Both are available for 34 of the 35 models.
The departures exceed $s_i$ in 33 of these and $t_i$ in all 34.

Figure~\ref{fig:ilq} shows the I-Love-Q relations, fitted in the same way to 37 models with $41.7\le\bar\Lambda\le873.5$.
The largest departures are $0.17\%$ for $\bar I$--$\bar\Lambda$, $0.33\%$ for $\bar Q$--$\bar\Lambda$ and $0.40\%$ for $\bar I$--$\bar Q$.
The corresponding RMS logarithmic residuals are $0.052\%$, $0.111\%$ and $0.131\%$.
These relations are thus roughly an order of magnitude tighter than $\bar I$--$C$, as for other EOSs~\cite{Yagi:2013iloveq,Yagi:2017approximate}.

We also refit the I-Love-Q relations with more freedom (not shown).
A separate offset for each $\theta$, with a common shape, lowers the largest departures to $0.10\%$, $0.16\%$ and $0.20\%$.
Separate fits for each $\theta$ lower them to about $0.03\%$.
A more flexible fit, however, also absorbs approximation error.
When one model at the end of a sequence is left out, the separate fits predict it with errors of up to $0.12\%$, $0.41\%$ and $0.37\%$.

These fits describe the five sequences of the present EOS family over the stated domains.
Table~\ref{tab:fits} lists the coefficients and degrees. The I-Love-Q fits use the same logarithmic polynomial as Eq.~\eqref{eq:fit}, with $X$ replacing $C$ and $Y$ replacing $\bar I$.
Full-precision coefficients, fit domains and numerical comparison details accompany the manuscript in the supporting files.
\begin{table}[t]
\caption{Coefficients of the logarithmic fits in Sec.~\ref{sec:results-relations}.}
\label{tab:fits}
\centering
\setlength{\tabcolsep}{3pt}
\begin{tabular}{lrrrr}
\hline\hline
$(X,Y)$ & $(C,\bar I)$ & $(\bar\Lambda,\bar I)$ & $(\bar\Lambda,\bar Q)$ & $(\bar Q,\bar I)$ \\
\hline
$n$ & 1 & 2 & 3 & 3 \\
$\mu$ & -1.5733 & 5.2513 & 5.2513 & 1.4793 \\
$s$ & 0.19074 & 1.5211 & 1.5211 & 0.46951 \\
$a_0$ & 2.3790 & 2.3333 & 1.4725 & 2.3384 \\
$a_1$ & -0.35782 & 0.36725 & 0.47917 & 0.36077 \\
$a_2$ & --- & 0.028492 & 0.0055323 & 0.024160 \\
$a_3$ & --- & --- & -0.0092402 & 0.0060488 \\
\hline\hline
\end{tabular}
\end{table}

\subsection{Observational implications}
\label{sec:results-obs}

Figure~\ref{fig:gw} compares the tidal deformabilities with GW170817.
We use the low-spin posterior samples of the LIGO--Virgo analysis~\cite{Abbott:2019properties} and draw the 50\% and 90\% regions estimated from a kernel density.
We adopt a source-frame chirp mass of $1.1858\,M_\odot$ at $z=0.0099$ and use Scott's bandwidth.
Each curve assumes that both stars follow the same $\theta$ sequence, at the source-frame chirp mass of the event.
All five curves lie inside the estimated 90\% credible region over their examined mass-ratio ranges.
Here $q$ is the lighter-to-heavier mass ratio.
With the baseline bandwidth, the curves for $\theta=0$, $\pi/3$ and $\pi/2$ also enter the 50\% region, for mass ratios $q\gtrsim0.68$--$0.70$.
The $\theta=2\pi/3$ curve enters it only for $0.75\lesssim q\lesssim0.99$, and not when the kernel bandwidth is halved.
The $\theta=\pi$ curve stays outside the 50\% region.

At the spin of PSR~J0740+6620, the rotating counterparts of the static maximum-mass configurations gain less than $1\%$ in mass (Table~\ref{tab:models}).
The $13\%$ change in $I$ at $1.338\,M_\odot$ quantifies the response to $\theta$ at fixed EOS parameters.

\FloatBarrier
\section{Conclusions}
\label{sec:conclusions}

We have investigated the rotational and static tidal responses of self-bound quark stars in general relativity using the axion-PNJL EOS of Ref.~\cite{Zhou:2026effects}.
The axion angle was treated as a fixed, spatially uniform parameter, and all remaining EOS parameters were held constant.
The calculation includes the finite-surface-density contributions to the rotational mass shift and the tidal response.

Increasing $\theta$ from $0$ to $\pi$ raises the static maximum mass from $1.677$ to $1.953\,M_\odot$.
These configurations remain below the lower edge of the quoted $1\sigma$ timing interval for PSR~J0740+6620.
At its spin frequency, the rotating counterparts of the static maximum-mass configurations gain $0.68$--$0.84\%$ in mass and $0.06$--$0.08$~km in equatorial circumferential radius.
The comparisons are made at fixed central energy density.

The dependence on $\theta$ is appreciable at fixed stellar mass.
At $M=1.338\,M_\odot$, the moment of inertia increases by $13\%$ and the dimensionless tidal deformability rises from $342$ to $547$ between the endpoint angles.
Over the common mass range examined, the differences in $I$ between adjacent sampled angles exceed the interpolation and table-resolution sensitivities tested here.
At fixed compactness, the Love number also varies systematically with $\theta$, with a spread that grows over the range examined.

The normalized quantities admit accurate joint fits across the five sequences.
Joint fits to the five sequences have maximum departures of $2.7\%$ for $\bar I$--$C$ and $0.4\%$ for the I-Love-Q relations over their fitted domains.
These fits describe the present EOS family over the stated domains.
The predicted binary tidal-deformability curves also lie within the estimated 90\% credible region of the adopted GW170817 posterior over the mass-ratio ranges examined, without establishing a constraint on $\theta$.

Future work could examine how these trends change when other EOS parameters are varied and extend the rotational analysis beyond the slow-rotation regime.

\section*{Data availability}
Data supporting the findings of this study are available from the corresponding author on reasonable request.

\par\vspace{\baselineskip}
\section*{Competing interests}
The author declares no competing interests. 

\section*{Funding}
This work was supported by the Deanship of Scientific Research, Vice Presidency for Graduate Studies and Scientific Research, King Faisal University, Saudi Arabia (Grant No. KFU266320).

\bibliography{references}

\end{document}